\documentclass[5p,12pt,authoryear]{elsarticle}

\title{Comic ray flux anisotropies caused by astrospheres}
\author[kls]{K.~Scherer\corref{cor1}}
\ead{klks@tp4.rub.de}
\author[dts]{R.D.~Strauss}
\ead{DuToit.Strauss@nwu.ac.za}
\author[dts]{S.E.S.~Ferreira}
\author[kls]{H.~Fichtner}
\address[kls]{Institut f\"ur Theoretische Physik IV, Ruhr-Universit\"at Bochum, Germany.}
\address[dts]{Centre for Space Research, North-West University, Potchefstroom, South Africa}
\cortext[cor1]{Corresponing author}
\usepackage{xspace}
\usepackage[switch]{lineno}
\usepackage{amsmath}
\newcommand{\degree}{\ensuremath{o}}
\newcommand{\s}{\ensuremath{\mathrm{S3}}\xspace}
\usepackage{twoopt,textcomp}
\newcommand{\permille}{\textperthousand}
\newcommand{\apjl}{Astrophys. J. Lett.}
\newcommand{\apj}{Astrophys. J. }
\newcommand{\ssr}{Space~Sci.~Rev.}%

\newcommand{\apss}{Astrophys. Space Sci.}
\newcommand{\aap}{Astron. \& Astrophys.}
\newcommand{\jgr}{J. Geophys. Res.}
\newcommand{\prd}{Phy. Rev. D}
\begin{document}

\begin{frontmatter}
\begin{abstract}
  Huge astrospheres or stellar wind bubbles influence the propagation
  of cosmic rays at energies up to the TeV range and can act as
  small-scale sinks decreasing the cosmic ray flux. We model such a
  sink (in 2D) by a sphere of radius 10\,pc embedded within a sphere
  of a radius of 1\,kpc. The cosmic ray flux is calculated by means of
  backward stochastic differential equations from an observer, which
  is located at $r_{0}$, to the outer boundary. It turns out that such
  small-scale sinks can influence the cosmic ray flux at the
  observer's location by a few permille (i.e\ a few 0.1\%), which is
  in the range of the observations by IceCube, Milagro and other large
  area telescopes.
\end{abstract}
\end{frontmatter}

\section{Introduction}

Large area cosmic ray detectors like the Tibet Air shower experiment,
IceCube/IceTop, Milagro and HAWC, among others observe a multipole
like anisotropy of the high energy cosmic ray flux (CRF) over the
entire sky
\citep{Iuppa-etal-2013,Abeysekara-etal-2014,BenZvi-2014,Desiati-2014,DiSciascio-Iuppa-2014,diSciascio-2015,Lopez-Barquero-etal-2015,Icecube-etal-2016}. The
energies of interest are in the TeV range which vary on small-scales
by a few permille (\permille) for details see \citet{Toscano-etal-2012} and
\citet{Iuppa-DiSciascio-2012}.

At higher energies (PeV) anisotropies were also found
\citep{Giacinit-Sigl-2012,Zotov-Kulikov-2012,Aartsen-etal-2013,Glushkov-Pravdin-2013}
which may be still of Galactic origin. Even at energies around EeV,
anisotropies in the arrival directions are obsereved
\citep[e.g.\ ][]{Abreu-etal-2011,Abreu-etal-2013}, which may be at the
transition to an extragalactic origin of the cosmic rays.
These high energies are not taken under consideration here.

A few explanations have been proposed, either related to interstellar
magnetic field variations \citep{Amenomori-etal-2011}, intermediate
turbulence \citep{Biermann-etal-2015} due to the heliotail
\citep{Desiati-Lazarian-2014,Zhang-etal-2014,Pogorelov-etal-2015,Schawdron-etal-2015}. A
detailed analysis of the power spectrum is discussed in
\citet{Ahlers-Mertsch-2015} where thet authors showed that the
strength of the power spectrum is related to the diffusion
tensor. \citet{Harari-etal-2016} discussed turbulent magnetic fields
as the cause of small angular scale variations, while
\citet{Battaner-etal-2015} correlated the anisotropy to the global
cosmic ray flux.

There are small-scale variations (tens of degrees) and even tiny-scale
variations about a degree or less in the interstellar medium caused by
astrospheres, planetary nebulae and similar inhomogeneities
\citep{Stanimirocic-etal-2010,Haverkorn-Spangler-2013,Linsky-Redfield-2014}. In
the following we explain how such variations -- due to the presence of
astrospheres -- act as small-scale sinks (\s) of CRF in the interstellar medium, can lead to such
anisotropies.

Huge astrospheres or stellar wind bubbles have been discussed for
example in \citet{Mackey-etal-2015} and \citet{Scherer-etal-2016},
especially their influence to the CRF was studied in
\citet{Scherer-etal-2015a} for the case of $\lambda$ Cephei. The
latter authors found that the CRF at energies up to 100\,TeV is
affected on scales below 1\,pc along the stagnation line of the
astrosphere. Because the discussed astrosphere of $\lambda$ Cephei is
special, in the sense that the relative motion between the star and
the ISM is high (about 80\,km/s), and the bow shock distance is about
1\,pc. Most of the astrospheres of hot stars do not show any relative
motion and build stellar wind bubbles of the order of
10-100\,pc. These bubbles have very high compression ratios
\citep{Toala-Arthur-2011} and thus can effectively act as sinks for
the CRF. The CRF is already affected directly beyond the bowshock, see
\citet{Scherer-etal-2011}, \citet{Strauss-etal-2013}, and
\citet{Luo-etal-2015}, and thus the effective modulation in
astrospheres starts directly behind the bow shock, and not as in the
helioshere at the heliopause \citep{Kota-Jokipii-2014,Potgieter-2014}.

Here we setup a model where we study the transport of cosmic rays
(CRs), when there is a \s between the outer boundary and the
observer. In section~\ref{sec:1} we present the model in detail, while
in section~\ref{sec:2} the numerical scheme is discussed. In
section~\ref{sec:3} we study the results for a range of appropriate
parameters (given in Table~\ref{tab:1} below) and finally we give a
r\'esum\'e in section~\ref{sec:4}

\section{The model}\label{sec:1}

\begin{figure}[t!]
  \centering
  \includegraphics[width=0.95\columnwidth]{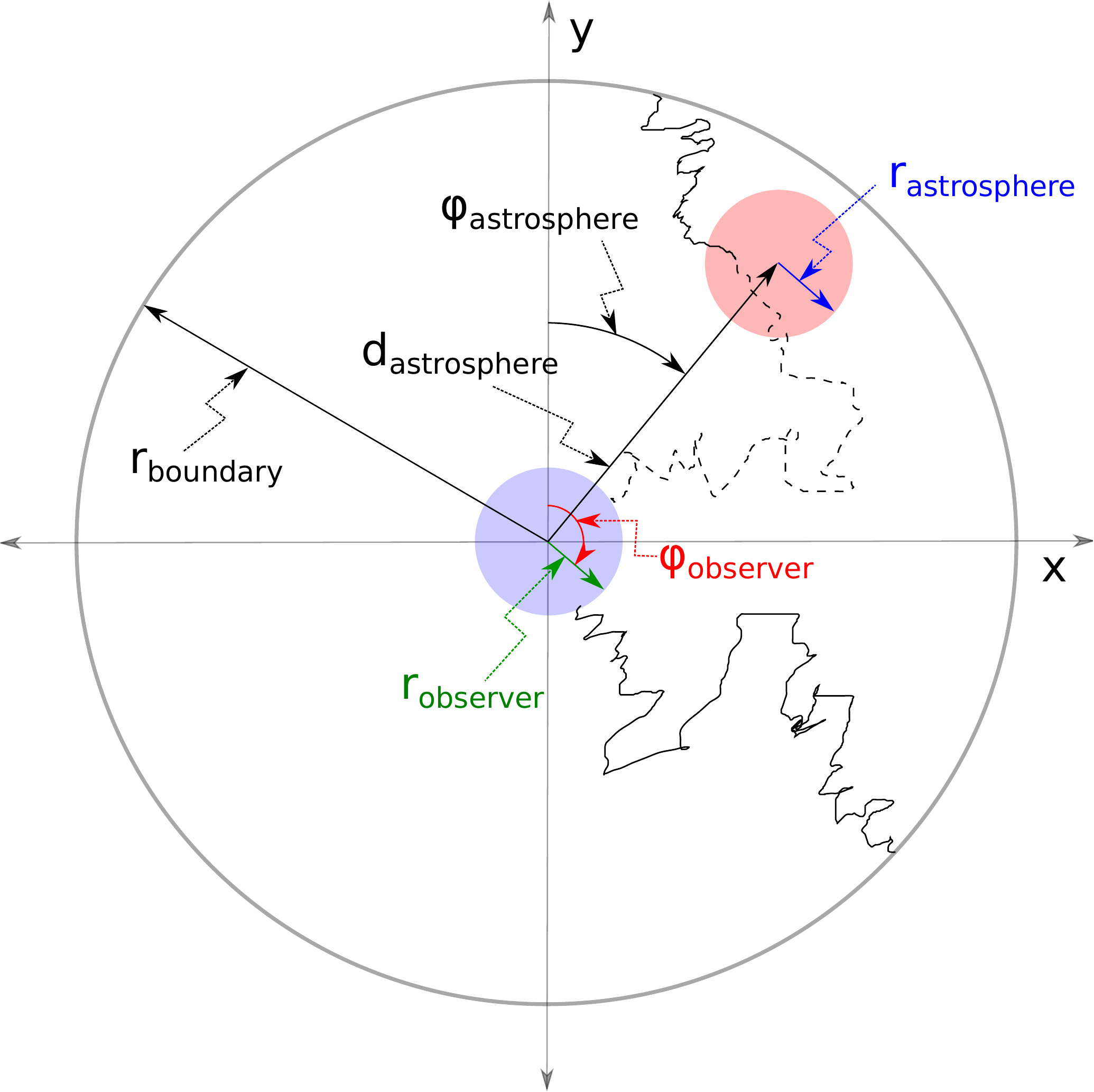}
  \caption{The model setup. The astrosphere (or \s) is located at a
    distance $d_{a}$ from the observer and has a radius of
    $r_{a}$. The whole integration area is a sphere with radius
    $r_{b}$.  The magnetic field is directed along the $y$-axis and
    the astrospheres is offset from the $y$-axis by an angle
    $\varphi_{a}$. The inner boundary is at $r_{o}$, which is much smaller
    than the CR mean free path. The two scatter lines show the path of
    a pseudo particles: one travels unhindered from $r_{b}$ to
    $r_{o}$, while the other interacts with the astrosphere and does no longer
    contribute (fully) to the CRF at $r_{o}$. The dotted line shows a
    possible path of that pseudo-particle.
}
  \label{fig:1}
\end{figure}

The basic scenario of our model is shown in Fig.~\ref{fig:1}: CRs,
propagate from their sources (specified at
$r_{\mathrm{boundary}}=r_b$) towards Earth, assumed to be at the
origin. We now place a spherical inhomogeneity (\s, most likely, an
astrosphere), with a radius of $r_{\mathrm{astrosphere}}=r_a$, at a
distance of $d_{\mathrm{astrosphere}}=d_a$ from the Sun. The position
of the astrosphere is also specified by the angle
$\varphi_{\mathrm{astrosphere}}=\varphi_a$, measured from the $y$-axis. We
do not calculate the intensity at Earth itself (this would be a
point in this set-up), but at an {\it observer's} position,
specified by $r_{\mathrm{observer}}=r_o$ and the angle
$\varphi_{\mathrm{observer}}=\varphi_o$. It is assumed that any anisotropies
at $r_o$ will be {\it frozen-in} and be directly observable at Earth;
a good approximation if the particle mean free path is
$\lambda > r_{o}$, which is usually larger in the ISM
\citet[for the local influence of different mean free pathes on
astrosphere, see ][]{Scherer-etal-2015a}. We assume that the \s
influence the particle intensity, i.e.\ some particles that interact
with the \s will be lost. This could be due to e.g.\ adiabatic energy
losses suffered in the astrosphere's expanding stellar wind or due to
catastrophic losses in a denser medium close the host
star. Independent of the process, we assume that the intensity of CRs,
when interacting with the astrosphere, will decrease.

The diffusion coefficient in the transport equation depend, in
general, on the magnetic field structure inside an astrosphere, which
is not known. Because we are not interested in the details of the CRF
inside the astrosphere, we simulate the CRF through it by an
extinction coefficient. For details see below.

As already mentioned, we require $\lambda > r_{o}$ for our simulation.
Furthermore also assuming $\lambda < r_{b}$, we can safely consider
that the resulting anisotropies will be small and use a Parker-like
transport equation \citep{Parker-1965} to describe their
transport in the turbulent interstellar medium (we therefore have the
scaling $r_{o} < \lambda < r_{b}$). For a nearly isotropic CR
distribution function $f$, we solve
\begin{equation}
\label{Eq:TPE}
\frac{\partial f}{\partial t} = \nabla \cdot \left(\mathcal{K} \cdot \nabla f \right) - \mathcal{Q}f
\end{equation}
which includes the loss-rate $\mathcal{Q}$ due to CRs interacting with
the \s. Assuming a 2D Cartesian geometry and a mean magnetic
field $\vec{B} = B \hat{y}$, the diffusion tensor reduces to
\begin{equation}
	\mathcal{K} =  \left[ \begin{array}{cc}
 \kappa_{\perp} & 0  \\
	0 & \kappa_{||}   \end{array} \right] 
\end{equation}
For simplicity, we assume $\kappa_{||} (=\kappa_y)$ and
$\kappa_{\perp}(=\kappa_x)$ to be constant and to be linearly related
via $\kappa_{\perp} = \eta \kappa_{||}$.

\section{The numerical solver}\label{sec:2}

We solve Eq. \ref{Eq:TPE} by means of stochastic differential
equations (SDEs). The set of SDEs, being equivalent to
Eq. \ref{Eq:TPE}, is, for $q \in \{ x,y \} $, simply
\begin{equation}
dq = d W_q \sqrt{2\kappa_q} 
\end{equation}
where $dW_q \approx \mathcal{R}_q \sqrt{\Delta s}$ are Wiener
processes and $\mathcal{R}_q$ independent, normally distributed,
random numbers. For details of this numerical approach
\citep[see e.g.\ ][]{ Strauss-etal-2011b,Kopp-etal-2012}. We solve these
equations backwards time, where backwards in time is labeled by $s$.

In general, a loss term is handled in the SDE formulation by keeping
track of the so-called {\it particle amplitude}, $\alpha_i$, for each
pseudo particle (labeled by $i$) during the integration
process. I.e., starting initially at $\alpha_i(s=0)=1$, this quantity
is updated as
\begin{equation}
\label{Eq:loss_rate}
\alpha_i(s+ \Delta s) = \alpha_i(s) \exp \left(- \mathcal{Q} \Delta s \right)
\end{equation}
We do, however, not yet know the precise form of the loss-rate,
neither the amount of time CRs will spend in such inhomogeneities
($\Delta s$, which will depend on the transport conditions in the
inhomogeneities themselves) so we approximate Eq. \ref{Eq:loss_rate}
as
\begin{equation}
\alpha_i(s+ \Delta s) \approx  \alpha_i(s) \left(1 -\chi   \right)
:{x,y \in \mathrm{\s }}.
\end{equation}
where we have defined an extinction coefficient $\chi$
($0\leq \chi \leq 1$), that is the fraction of the CR particles lost
when a \s was encountered, i.e.\ $\chi = 1$, or equivalently
$\chi = 100\%$, means that all particles that encountered the
\s were lost.

The numerical time step in the SDE scheme, $\Delta s$, determines what
scale of structures will be sampled by these pseudo-particles in
configuration space. In order to sample the relevant spatial features
in our model set-up (i.e.\ pseudo-particles should not {\it jump}
across the \s, but sample it continuously), we follow
\citet{Strauss-etal-2013} and implement a variable time step as
\begin{equation}
\Delta s = \sigma  \frac{\mathcal{L}^2}{\max \left\{\kappa_{||}, \kappa_{\perp}  \right\} }  
\end{equation}
where $\mathcal{L}$ is the spatial extend of the smallest structure in
the model and $\sigma$ a constant, usually chosen to be
$ \sigma = 0.5$ or smaller.

The solution of the transport equation at any point in space is now
calculated, in the steady-state limit, as
\begin{equation}
\label{Eq:error_bar}
f(x,y) \approx \frac{1}{N}  \sum_{i=1}^{N} \alpha_i f_{\mathrm{boundary}} 
\end{equation}
where $N$ is the number of pseudo-particles (a term for each numerical
realisation of the set of SDEs, not a physical CR particle) solved and
$f_{\mathrm{boundary}}$ is the boundary value, specified at $r_b$. We simply use
$f_{\mathrm{boundary}}=1$ without any loss of generality. 

We can easily estimate the statistical error related to this method of
solution by assuming Poisson statistics for the resulting distribution
(a fair approximation when looking at the results of e.g.\
\cite{Strauss-etal-2011a}), $N = N \pm \sqrt{N}$, so that
Eq.~\ref{Eq:error_bar} is modified to become
\begin{equation}
f(x,y) \approx \frac{1}{N}  \sum_{i=1}^{N} \alpha_i  \left(1 \pm \frac{1}{\sqrt{N}}  \right),
\end{equation}
and the resulting {\it error bar} plotted with the intensities in the
next sections. To get our results accurate to the permille range, we
therefore need, at least, $N = 10^6$, pseudo-particles so, for the
simulation in this work, we use $N = 10^7$ such particles at each
phase-space position where the intensity is calculated.

\section{Results}\label{sec:3}

Because we study a large set of parameters such as the variation of the
distance $d_{a}$ to the astrosphere, its size $r_{a}$, the angle to
the observe $\varphi_{a}$ as well the ratio $\eta$ of the perpendicular
to the parallel diffusion coefficient, and finally the loss rate
$\chi$ inside the astrosphere, we start with a reference solution (RS)
with $d_{a}=600$\,pc, $r_{a}=10$\,pc, $\varphi_{a}=45^{\degree}$,
$\eta=0.5$, and $\chi=100$\%.  Other model parameters relative
to the RS is shown in Table~\ref{tab:1}.

\begin{figure}[!ht]
    \centering
    \includegraphics[width=0.95\columnwidth]{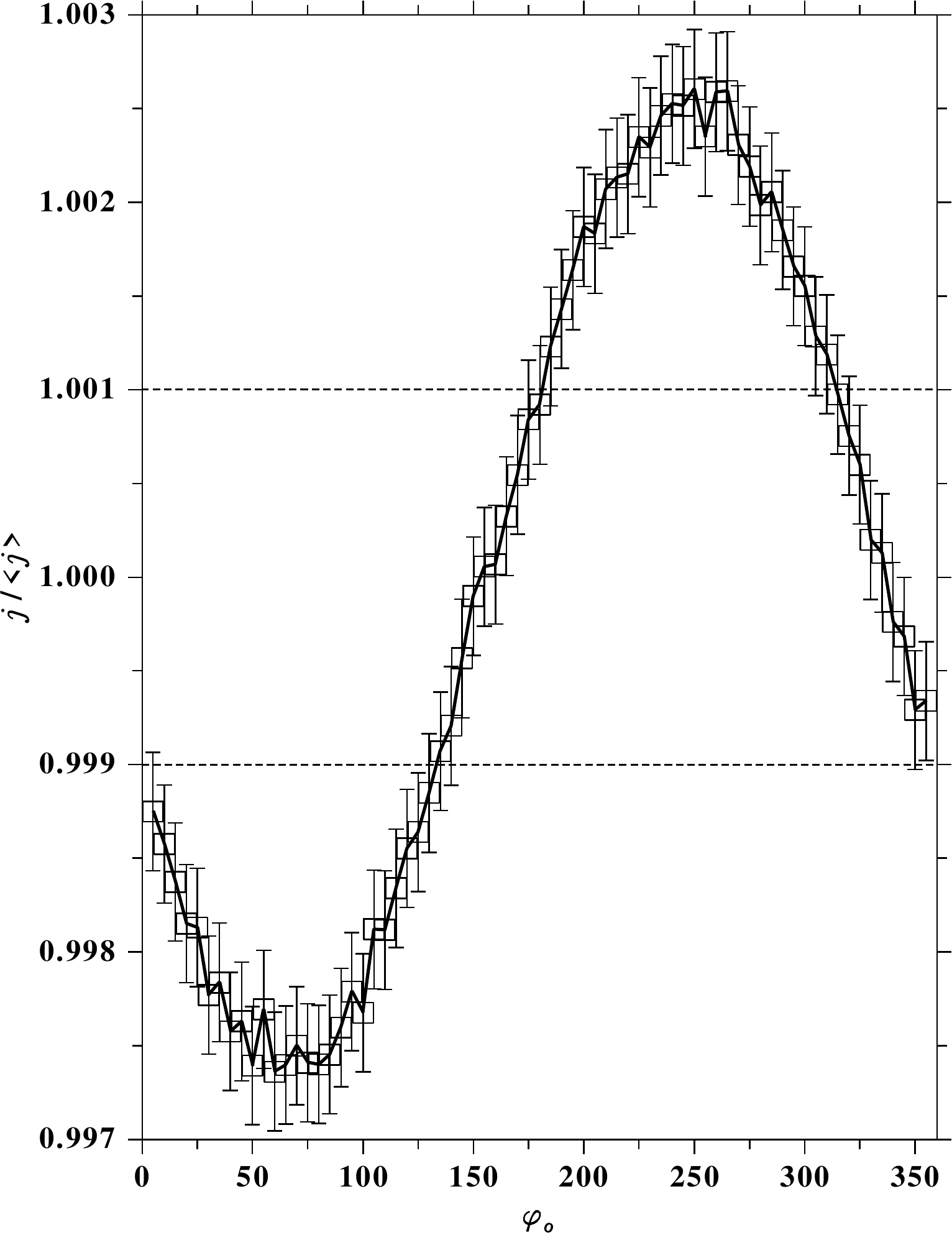}
    \caption{Reference solution for the CRF. To guide the eye the red
      lines show the 0.1\% level. As reference values we take a
      distance to the \s of $600$\,pc, a radius of $10$\,pc,
      an angle from the y-axis of $\varphi=45^{o}$, a ratio of
      $\eta=\kappa_{\perp}/\kappa_{\parallel}=0.5$. and the loss of the
    CRF flux inside the \s $\chi=100$\%. The variation of the
    CRF is about $\pm 0.3\%$ along the observer angle $\varphi$. Because
    the magnetic field is directed along the y-axis, the minimum is
    shifted by about 25\% from the actual position, in terms of
    $\varphi_{o}$ of \s.}
    \label{fig:2}
\end{figure}

\begin{figure*}[!ht]
    \centering
    \includegraphics[width=0.3\textwidth]{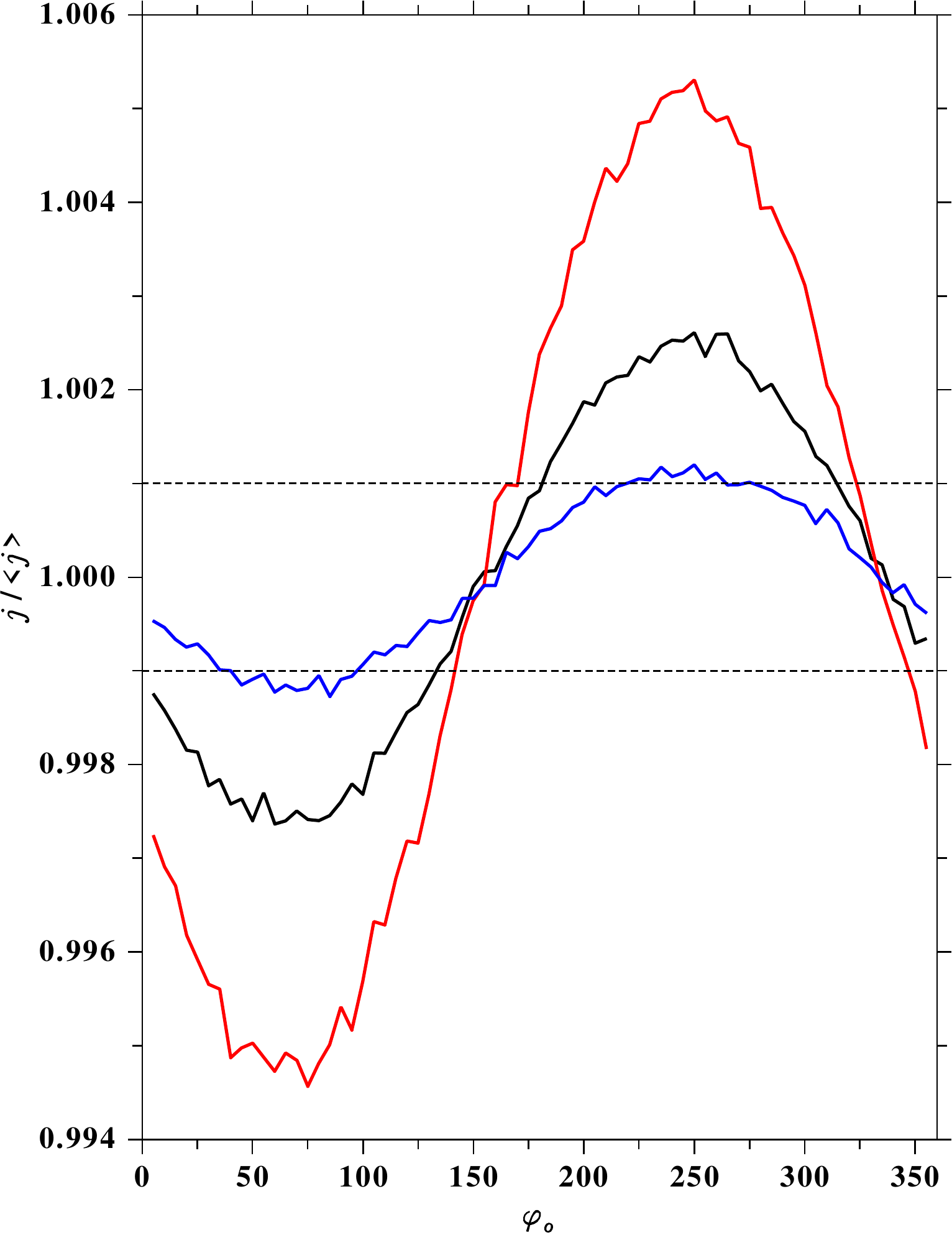}
    \includegraphics[width=0.3\textwidth]{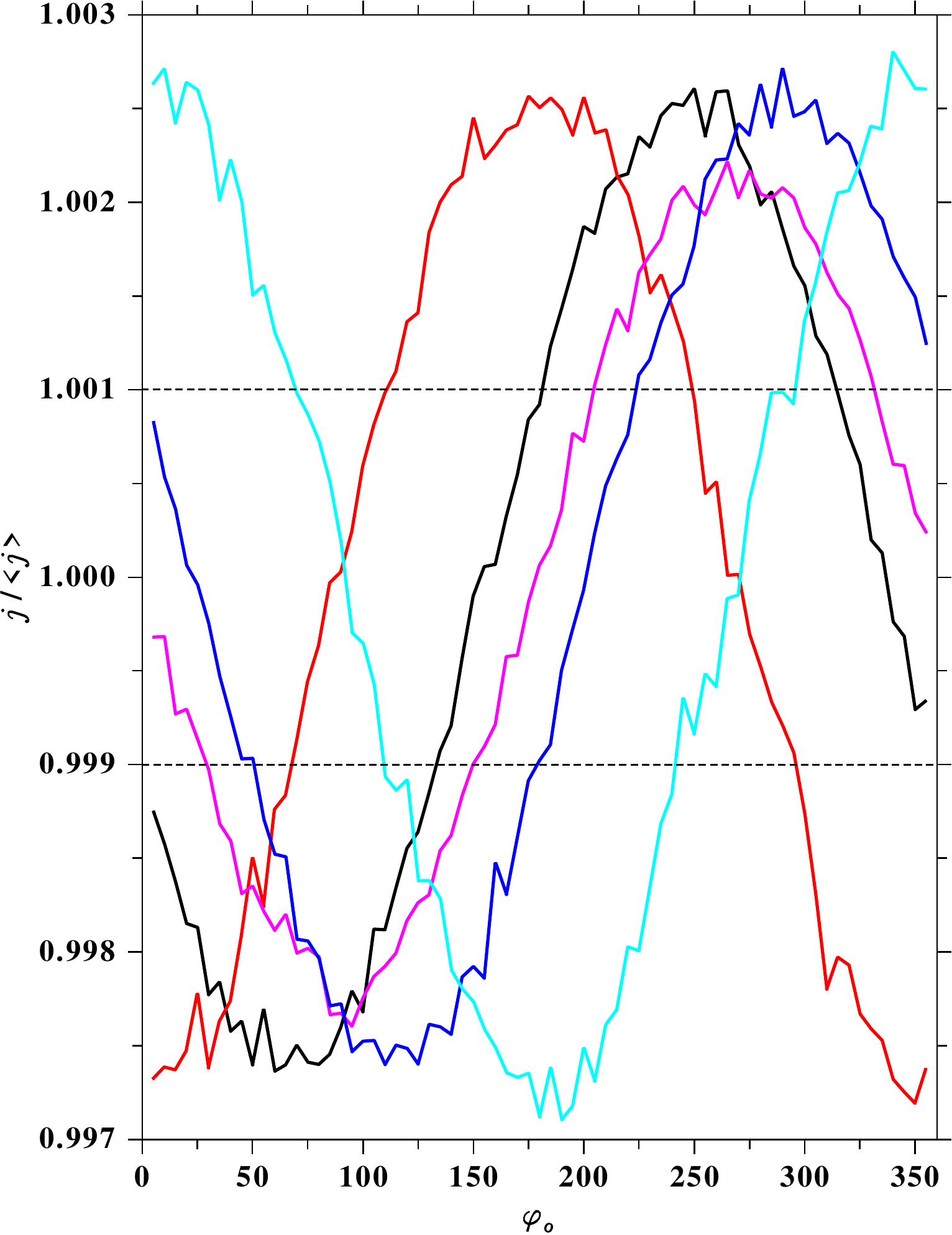}
    \includegraphics[width=0.3\textwidth]{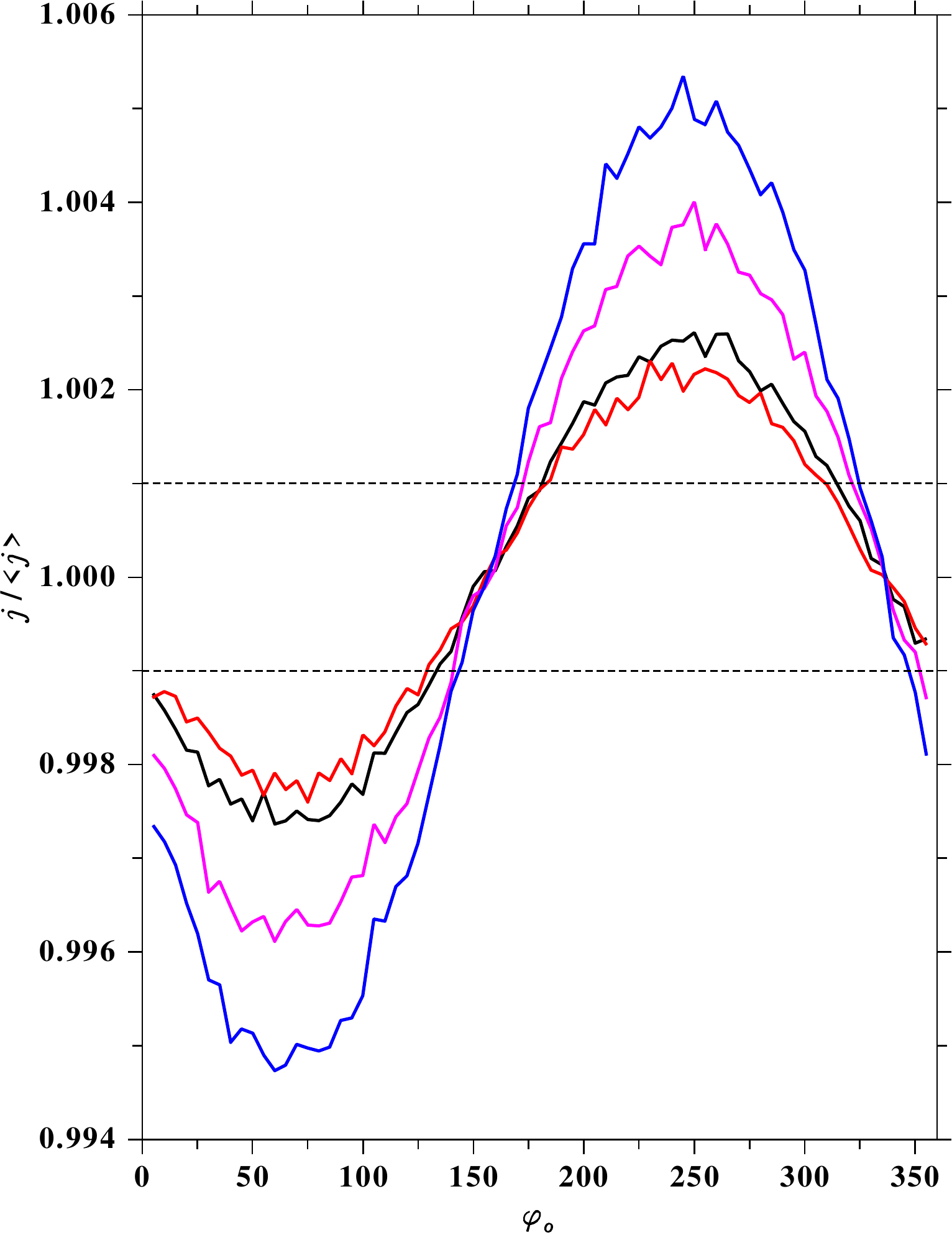}
    \caption{Variation of astrosphere distance (left panel), the angle
      with respect to the magnetic field $\varphi_{a}$ (middle panel),
      and the size of \s $r_{a}$ (right panel). The colour coding in
      the left panel is: $d_a=400$\,pc red line 
      and $800$\,pc blue line. The colours in the middle panel vary for
      $\varphi_{a}=0,45,90,135,180$\,degree as red, black, magenta, blue,
      and cyan, respectively. Finally those in the right panel, where
      the size varies for $r_a=5,10,25,50$\,pc, as red, black,
      magenta, and blue. The black lines always indicate the RS. The
      scales on the $y$-axis differ.}
    \label{fig:3}
\end{figure*}

\begin{figure*}[!ht]
    \centering
    \includegraphics[width=0.3\textwidth]{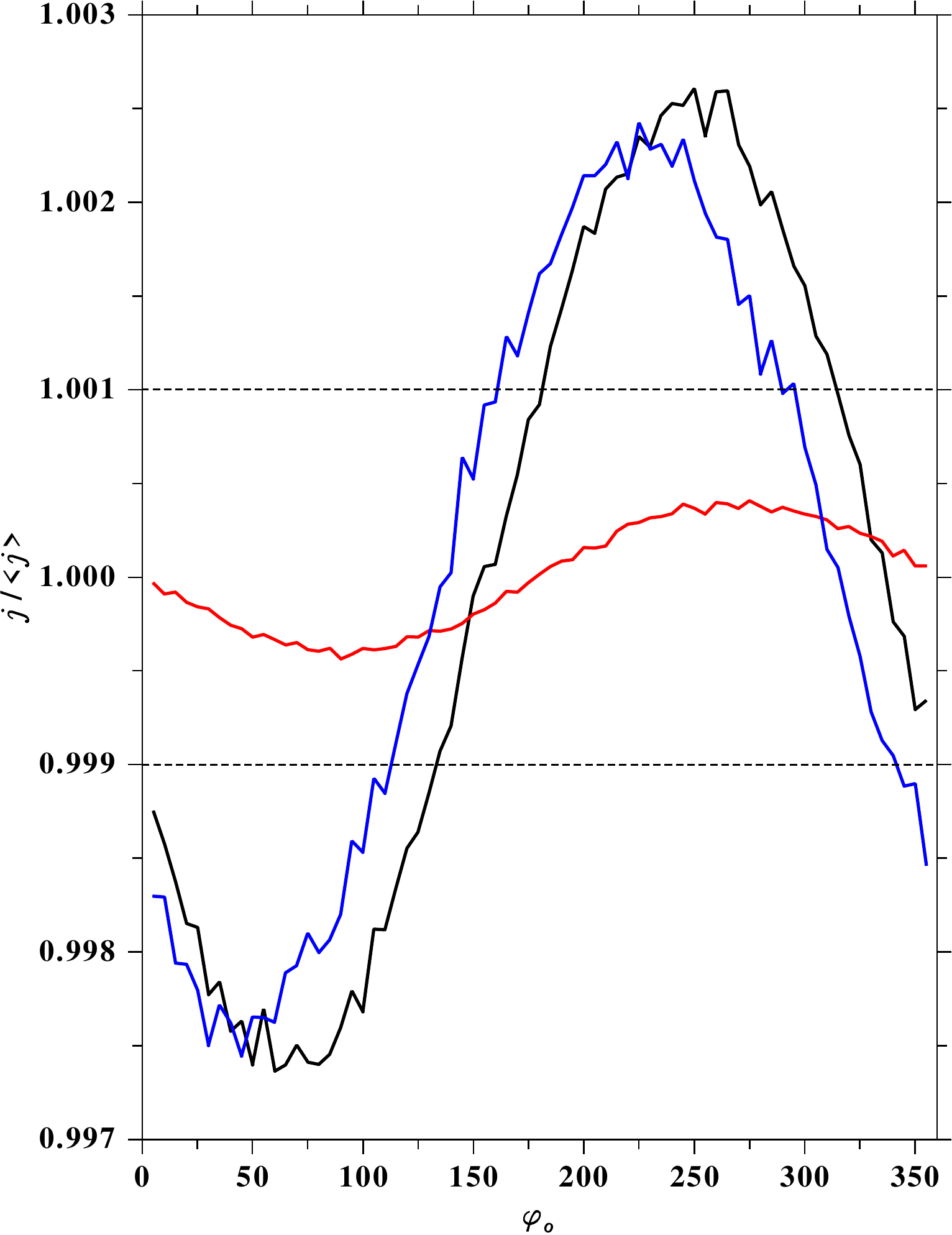}
    \includegraphics[width=0.3\textwidth]{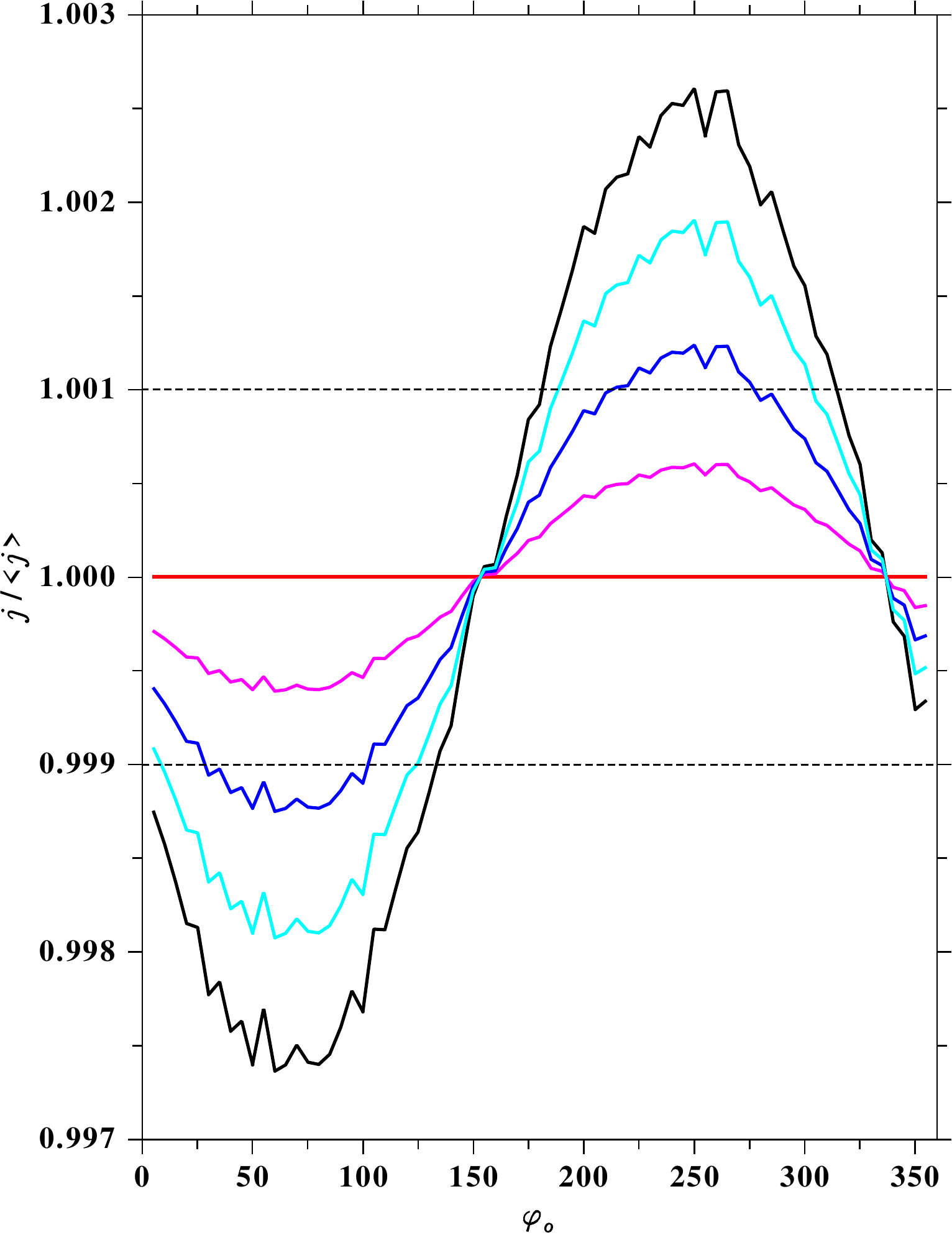}
    \includegraphics[width=0.3\textwidth]{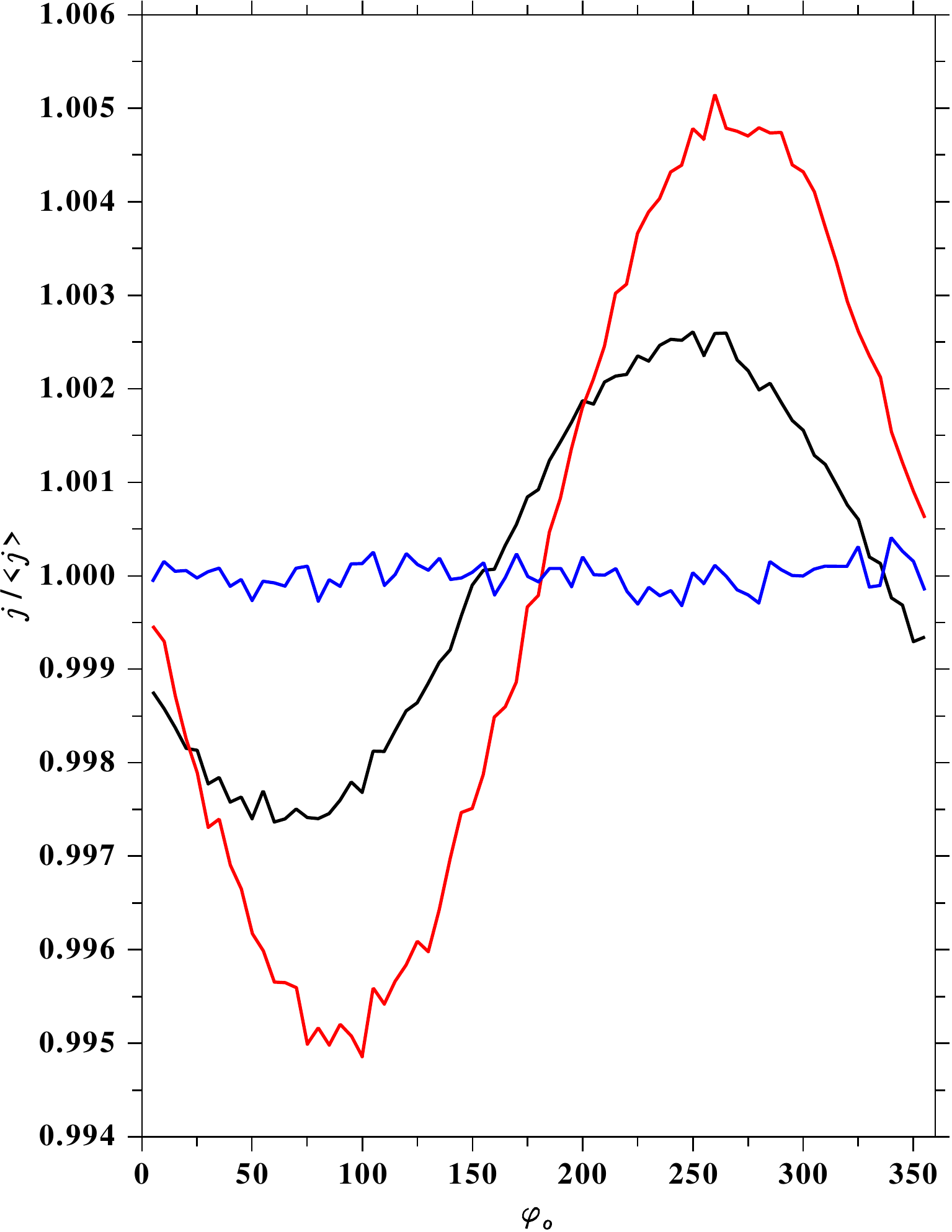}
    \caption{Variation of the ratio $\eta$ of perpendicular to
      parallel diffusion coefficients (left panel), the extinction
      coefficient $\chi$ (middle panel) and the effect of two \s
      (right panel). The colours for $\eta=0.02,0.5,1$ are red, black,
      blue (left panel), that of $\chi=0,25,50,75,100$, red, magenta,
      blue, cyan, black. Again the black line represents the RS. In
      the right panel we have modeled two astrospheres separated by
      90$^{\degree}$ and 180$^{\degree}$. The latter normalised flux
      does not show any variation, because of the ``destructive''
      interference, while the former has an ``constructive''
      interference and an additional shift of the maxima/minima.}
\label{fig:4}
\end{figure*}

The variation of the CRF of the RS is shown in Fig.~\ref{fig:2}
together with the numerical error bars (see Eq.~\ref{Eq:error_bar}). The error
bars are similar in all subsequent models and not shown further. In
Fig.~\ref{fig:3} to~\ref{fig:4} the RS is, as reference, always shown
as the black line.

The RS shown in Fig. \ref{fig:2} is the solution of the SDE equation,
as a function of $\varphi_{o}$, at $r_{o} = 10$ pc. The solution is
normalised to its average value (averaged over $\varphi_o$), that is,
$j/\langle j \rangle > 1$ shows an excess of particles
{\it with respect to the average} of $j$ over all $\varphi_{o}$'s. Note
that, in the current model, the differential intensity $j$, is
directly proportional to the distribution function $f$. The black
lines horizontal show the permille range.

The solutions depend on the ratios
$\eta = \kappa_{\perp}/\kappa_{\parallel}$ and $\eta/\chi$. This is
because, for a diffusion equation, with constant (time independent)
coefficients and boundary conditions, the steady state solution is
only determined by the boundary conditions. When, in future, we
generalize Eq. \ref{Eq:TPE} to include energy losses in the
interstellar medium, and implement more realistic parametrisations of
the energy losses in \s, the magnitude of
$\kappa_{\parallel},\kappa_{\perp}$ may play an important role (here
the ratio $\kappa/Q$ plays that role).

\begin{table}[t!]
  \centering
  \begin{tabular}[t!]{llllll|ll}
  \hline
     \# &   $r_{a}$  & $d_{a}$  & $\varphi_{a}$ & $\eta$ & $\chi$ & A & $\beta$    \\
        &  pc & pc & $^{o}$ & & \% & \permille & $^{o}$\\
          \hline
          \hline
    0        & 10  & 600 &  45 & 0.5 & 100  & -2.6  & - 70 \\
    \hline
    1        & 10  & 400 &  45 & 0.5 & 100  & -5.2  & - 65 \\
    2        & 10  & 800 &  45 & 0.5 & 100  & -1.1  & - 70 \\
    \hline
    3        & 10  & 600 &    0 & 0.5 & 100  & -2.6 & -   0 \\
    4        & 10  & 600 &   90 & 0.5 & 100  & -2.3 & -  90 \\
    5        & 10  & 600 &  135 & 0.5 & 100  & -2.6 & - 110 \\
    6        & 10  & 600 &  180 & 0.5 & 100  & -2.7 & - 180 \\
    \hline
    7        &  5  & 600 &   45 & 0.5 & 100  & -2.2 & - 70\\
    8        & 25  & 600 &   45 & 0.5 & 100  & -3.8 & - 70\\
    9        & 50  & 600 &   45 & 0.5 & 100  & -0.5  & - 70\\
    \hline
   10        & 10  & 600 & 45  & 0.02 & 100  & -0.4 & - 90\\
   11        & 10  & 600 & 45  & 1.0  & 100  & -2.4 & - 45\\
   \hline
   12        & 10  & 600 & 45  & 0.5  &   0  & 0      & --   \\
   13        & 10  & 600 & 45  & 0.5  &  25  & -0.6 & - 70\\
   14        & 10  & 600 & 45  & 0.5  &  50  & -1.2 & - 70\\
   15        & 10  & 600 & 45  & 0.5  &  75  & -1.8 & - 70\\
   \hline
    16        & 10  & 600 &  45/135 & 0.5 & 100  & -5.1  & - 90 \\
    17        & 10  & 600 &  45/225 & 0.5 & 100  & 0 & -- \\
   \hline
  \end{tabular}
  \caption{The parameters used for the different models. Model 0 is
    the RS and each vertical line separates a new set of parameters:
    models~1 and~2 vary the distance, models~3 to~6 the position
    angle, i.e.\ the magnetic field direction with respect to the \s,
    model~7 to~9 the size of the \s, model~10 and~11 the the ratio
    $\eta$ between the perpendicular to parallel diffusion, and
    models~12 to~15 the extinction coefficient $\chi$. The last two
    columns are the parameters of a fit to $-A \cos(\varphi-\beta)$ (see
    section~\ref{summ} below). In models~16 and~17 two astrospheres
    are modeled simultaneously, which are identically to model~0,
    except is position angle.} 
  \label{tab:1}
\end{table}

The RS show some interesting features: The variation is in the permille
range as required by the observations. The most exciting fact is, that
the \s, with a filling factor
$F = r_{a}^{2}/r_{b}^{2} \approx 3\cdot10^{-4}$ (defined as the ratio
of the area of the \s to that of the large sphere, see section~\ref{summ} below), influences the
variation along the observer angle affects over the entire range,
with a maximum at $\approx 250^{o}$, which is the undisturbed CRF
flux, and a minimum at $\varphi_{\mathrm{min}}\approx 70^{o}$. The latter
has an offset $\Delta \varphi = \varphi_{\mathrm{min}}-\varphi_{o}$ of about
$25^{o}$ to the direction of the \s. This is caused by the orientation
of the magnetic field along the $y$-axis and the fact, that the cosmic
rays are not only diffusing along the magnetic field but also
perpendicular to it. In our setup of the model (Fig.~\ref{fig:1})
two possible paths of a pseudo-particle are indicated: one which is not
affected by the \s (solid line) and another one which is (solid and
then dashed line). The pseudo-particle would follow the dashed line if
it is not absorbed in the \s. This is the case when we change the
extinction coefficient $\chi$ for some particles.

Another exciting feature is the extent of the minimum (maximum) of the
CRF flux, it can be as broad as some ten degrees. This is also
compatible within the range of the observed angular size of the CRF
\citep[see, e.g.\ ][]{Icecube-etal-2016}.

To study these effects further we varied first the distance of the \s
to $d_{a}= 400$\,pc, which can be seen in the left panel of
Fig.~\ref{fig:1} (model 1, red line) and to $d_{a}=800$\,pc (model 2,
blue line). The amplitude of the curves increases (decrease) with
decreasing (increasing) distance compared to the RS, but their minima
and maxima remain more or less at the same position. The increase in
the amplitude is not linear: it is about a factor 2 larger for model~2
and about a factor 0.4 smaller for model~1 compared to RS. The reason
is that a closer \s blocks relatively more CR because of its apparent
large area with respect to the observer.  These effects are expected:
the further away a \s is, the smaller is the CRF variation.

In the middle panel we show the variation in the position angle
$\varphi_{a}$. This changes the orientation of the magnetic field
relative to \s, and thus the transport of CR along the magnetic field
is reduced as can be seen by the magenta line, where the propagation
from the \s to the observer is perpendicular to the magnetic field and
thus the amplitude of the CRF is slightly reduced. The parallel and
anti-parallel propagation effects shown by the red and cyan lines seem
to be marginally higher than the RS, but that is inside the error
bars. Nevertheless, the minima (maxima) of the models with
non-parallel propagation (black and blue lines) deviate from the
location of the position angle by $\Delta \varphi \approx 25^{o}$ for
$\varphi_{a}=45^{\degree}$ and $\Delta \varphi\approx -25^{o}$ for
$\varphi_{a}=135^{\degree}$, while the model the perpendicular propagation
has its minimum at $\Delta \varphi\approx 0^{\degree}$. Thus a varying
magnetic field orientation influences the position of the CRF-minima
(maxima) from $0^{\degree}$ up to $\approx 25^{\degree}$. Thus determining
the position of the sinks by observations requires a good knowledge of the
interstellar magnetic field.

Finally, we varied the ratio $\eta$ of the perpendicular to parallel
diffusion coefficient and the extinction coefficient $\chi$ shown in
Fig.~\ref{fig:4}. The variation of $\eta$ is shown in the left panel
and that of $\chi$ in the middle panel. It can be seen that for very
small ratios $\chi$ the amplitude of the CRF is below 0.1\% (red line,
right panel) and that the minima (maxima) are shifted to much larger
offsets than changing the position angle. If both diffusion
coefficient are equal ($\kappa_{\parallel}=\kappa_{\perp}$) the
extrema of the CRF are a little smaller than that of the RS and are
not offset from the position angle $\Delta \varphi \approx
0^{\degree}$. The latter is due to the fact, that if $\eta=1$ both
diffusion coefficients are equal and thus there is no preferred
direction for diffusion (isotropic or scalar  diffusion).

In the middle panel of Fig.~\ref{fig:4} the extinction coefficient is
changed. This varies only the amplitude from zero for no extinction
($\chi=0\%$) to that of our RS. An extinction coefficient of $\chi=0$
or $\chi=25\%$ are below the 0.1\% level, while that of
$\chi=50\%,75\%$ and $\chi=100\%$ lead to variations of the CRF flux
in the observed range (for a discussion see below section \ref{summ}).

We just shortly present the case (model~16) if we have two \s locate at
$\varphi_{a,1}=45^{\degree}$ and $\varphi_{a,2}=225^{\degree}$ (see
Fig:~\ref{fig:4}, right panel). Because the
diffusion depends on the direction and not on the orientation of the
magnetic field, it is identical for both \s and thus where the CRF
of one \s shows a maximum, the other \s has a minimum, and analogously
 $\varphi_{o}$ both variations add to zero. Thus the result is like
that of a ``destructive'' interference. The absolute CRF should be
lower compared to a single \s, but the absolute flux cannot be studied
with the present setup of our model.  Another case, where the astrospheres are
located at $\varphi_{a,1}=45^{\degree}$ and $\varphi_{a,2}=135^{\degree}$ (see
Fig:~\ref{fig:4}, right panel) gives an ``constructive'' interference
and an additional phaseshift compared to the RS. We presented these two
cases here to show the capability of our model, but do not discuss it
further in the present context. Nevertheless, the above examples show, that we
can get a kind of interference of different sinks, which can be
decomposed into a multipole power spectrum
\citep{Ahlers-Mertsch-2015}.

\subsection{Summary of the parameter study}
\label{summ}
 We fitted the function
$-A_{i} \cos(\varphi - \beta_{i})$ to each CRF in
Figs.~\ref{fig:2} to~\ref{fig:4} (not shown), where $i$ indicates the
model number. With the help of that function, it is easier to discuss
the amplitude $A_{i}$ and phase shift $\alpha_{i}$. The values of
$A_{i}$ and $\alpha_{i}$ are given in the last two rows of
Table~\ref{tab:1}.

We can identify two geometrical effects: the distance to the \s and
its radius: The amplitudes $A_{i}$ of the normalised CRF varies with
both the distance and the size of the \s (models~1 and~2, see left
panel and models ~7 to~9 right panel of Fig.~\ref{fig:3}). The
variations in the CRF amplitude are as expected: larger (smaller) for
smaller (larger) distances or sizes. Our expectation is that the
amplitudes depent on the opening angle of the \s as seen from the
observer. But because of the the few examples it is not possible to
study these effect empirically. 

The physics of the transport of CRs are influenced by the extinction
parameter $\chi$, which already gives amplitudes in the permille
range, when it is larger than 25\%, as can be expected for \s's
(models~12 ~14, middle panel of Fig.~\ref{fig:4}). This effect is also
as expected, the larger the extinction coefficient the larger is the
amplitude.

The ratio of the perpendicular to parallel diffusion coefficient is a
parameter, which is important not only for the amplitude but also for
the offset $\Delta \varphi$ (models~10 and ~11, right panel of
Fig.~\ref{fig:4}). This can be understood as an effect of the
parallel/perpendicular diffusion, and, thus, how easy CRs can travel
along the respective direction, influencing the amplitude and the
offset.  Thus, it again turns out that the perpendicular diffusion
coefficient plays an important role when modeling the galactic cosmic
ray transport, as already discussed in e.g.\
\citet{Effenberger-etal-2012}, \citet{Kumar-Eicheler-2014}, and
\citet{Mertsch-Funk-2015}.

Finally, the orientation of the magnetic field plays an important role
affecting the offset $\Delta \varphi$. This is simulated by the
position of the \s relative to the x-axis (model~3 to~6 middle panel
of Fig.~\ref{fig:3}). This can be explained by the fact that there is a more
efficient diffusion in the parallel direction, as long as
$\eta \ne 1$, which then causes the offset. 

All the models discussed here show a relatively flat minimum (maximum)
with an extension of a few degrees, where the normalised CRF
variation are inside the error bars (only shown in Fig.~\ref{fig:2} for
the RS). Thus, from an observational point of view the sources can
have a large angular extent.

Thus we have five parameters ($d_{a},r_{a},\varphi_{a},\eta$, and $\chi$)
which influence the amplitude and offset of the CRF. We have varied
them individually with respect to the RS, but did not study
simultaneous variations of these parameters. This can be done when
observational data are available. Here we demonstrated the effects to
the CRF of a \s with a small filling factor of $F=10^{-4}$, and
conclude that such a \s is a possible explanation for the observed
small-scale variations. 

Replacing the sinks by appropriate sources we anticipate to get
inverse results, because at the source position we expect to have the
largest flux going to a minimum opposite to it. The phase shift should
also be equal to that of the sinks discussed here.We will study it in
a forthcoming paper.

We studied additionally, the CRF when two \s are present
(model~16 and~17, right panel of Fig.~\ref{fig:4}) which show something like
an interference. A detailed study of two or more \s would go far
beyond the scope of this work, but the two examples indicate that a
Fourier decomposition of the signal along the observer angle can give
some hints of the involved sinks.

In Eq.~\ref{Eq:TPE} we used a constant parallel diffusion
coefficient $\kappa_{\parallel}$ and loss term $\mathcal{Q}$, and,
thus get an ``energy dependence'' by increasing (decreasing)
$\kappa_{\parallel}$ relative to $\mathcal{Q}$, because of the
quasi-stationarity of Eq.~\ref{Eq:TPE}. For more realistic scenarios
one can include an energy dependent diffusion, but for the higher energies this
does not play an essential role.  

The filling factor $F$ is different when we go from a 2D scenario to a
3D one. Then one needs a sphere with an radius of approximately 50\,pc
to get the same filling factor. Also the diffusion perpendicular to
the galactic plane can influence our estimates. Assuming that cosmic
rays diffusing perpendicular to the galactic disk escape before they
can be detected and the CRF is confined in the galactic disk, then the
modulation is approximately 2D and we expect an similar result.

In the above models we used, except for models~12 to~15 a extinction
rate of 100\%, which may be too large and one would expect it to be in
range 25\% to 50\%. But this is only a sophisticated guess, because as
far as we know there is no model in which an astrosphere or other
object blocks (partially) the CRF. Only \citet{Scherer-etal-2015a} have
calculated the CRF along the stagnation line into the astrosphere
around $\lambda$ Cephei up to an inner boundary of 0.03\,pc.  Thus, to
get a better estimate on the extinction factor of the CRF, we need to
extend our calculations of the modulation to a point outside of the
astrosphere.

\section{Resume}\label{sec:4}

We simulated the cosmic ray flux in a 1\,kpc sphere, when 
small-scale sinks are located inside this sphere. We showed that there
is a depletion of the CRF in the order of a few permille. Moreover, it
turned out that such a small obstacle, for example, an astrosphere
with a radius of 10\,pc influences the entire observer angle, and is
not only a spike, but a sinusoidal variation. The flat minimum of
these variations have an extension of a few degree.

Thus, obstacles with filling factors of $F\approx10^{-4}$ lead to the
observed variations in the permille range, except when the ratio of
the diffusion coefficients ($\eta < 0.02$) or the extinction
coefficient ($\chi<25\%$) is too small. But then an increase in the
size or distance to the \s can compensate the low CRF variation.

We have demonstrated, that small-scale sinks in the ISM can lead to the
observed multipole character of the CRF. Unfortunately, due to the fact
that the arrival direction of the CRF in the minimum does not necessarily
coincide to with the location of the obstacle, the identification of the
latter is complicated.

We have shown that in the above simplified scenario, we obtain quite
promising results in describing the CRF anisotropy. Thus, we will
continue the study in future implementing more realistic scenarios.

\section*{Acknowledgement}
KS is grateful to the
\emph{Deut\-sche For\-schungs\-ge\-mein\-schaft, DFG\/} for funding
the projects SCHE 334/9-1 and SCHE334/9-2. This work is
based on research supported in part by the National Research
Foundation (NRF) of South Africa. This work was carried out within the
framework of the bilateral BMBF-NRF project ``Astrohel'' (01DG 15009)
funded by the Bundesministerium f\"ur Bildung und Forschung. The
responsibility of the content of this work is with the authors. 


\begin{thebibliography}{45}
\expandafter\ifx\csname natexlab\endcsname\relax\def\natexlab#1{#1}\fi
\expandafter\ifx\csname url\endcsname\relax
  \def\url#1{\texttt{#1}}\fi
\expandafter\ifx\csname urlprefix\endcsname\relax\def\urlprefix{URL }\fi

\bibitem[{{Aartsen} et~al.(2013){Aartsen}, {Abbasi}, {Abdou}, {Ackermann},
  {Adams}, {Aguilar}, {Ahlers}, {Altmann}, {Andeen}, {Auffenberg}, and
  et~al.}]{Aartsen-etal-2013}
{Aartsen}, M.~G., {Abbasi}, R., {Abdou}, Y., {Ackermann}, M., {Adams}, J.,
  {Aguilar}, J.~A., {Ahlers}, M., {Altmann}, D., {Andeen}, K., {Auffenberg},
  J., et~al., Mar. 2013. {Observation of Cosmic-Ray Anisotropy with the IceTop
  Air Shower Array}. \apj 765, 55.

\bibitem[{{Abeysekara} et~al.(2014){Abeysekara}, {Alfaro}, {Alvarez},
  {{\'A}lvarez}, {Arceo}, {Arteaga-Vel{\'a}zquez}, {Ayala Solares}, {Barber},
  {Baughman}, {Bautista-Elivar}, {Belmont}, {BenZvi}, {Berley}, {Bonilla
  Rosales}, {Braun}, {Caballero-Mora}, {Carrami{\~n}ana}, {Castillo}, {Cotti},
  {Cotzomi}, {de la Fuente}, {De Le{\'o}n}, {DeYoung}, {Diaz Hernandez},
  {D{\'{\i}}az-V{\'e}lez}, {Dingus}, {DuVernois}, {Ellsworth}, {Fiorino},
  {Fraija}, {Galindo}, {Garfias}, {Gonz{\'a}lez}, {Goodman}, {Gussert},
  {Hampel-Arias}, {Harding}, {H{\"u}ntemeyer}, {Hui}, {Imran}, {Iriarte},
  {Karn}, {Kieda}, {Kunde}, {Lara}, {Lauer}, {Lee}, {Lennarz}, {Le{\'o}n
  Vargas}, {Linnemann}, {Longo}, {Luna-Garc{\'{\i}}a}, {Malone}, {Marinelli},
  {Marinelli}, {Martinez}, {Martinez}, {Mart{\'{\i}}nez-Castro}, {Matthews},
  {McEnery}, {Mendoza Torres}, {Miranda-Romagnoli}, {Moreno}, {Mostaf{\'a}},
  {Nellen}, {Newbold}, {Noriega-Papaqui}, {Oceguera-Becerra}, {Patricelli},
  {Pelayo}, {P{\'e}rez-P{\'e}rez}, {Pretz}, {Rivi{\`e}re}, {Rosa-Gonz{\'a}lez},
  {Ruiz-Velasco}, {Ryan}, {Salazar}, {Salesa Greus}, {Sandoval}, {Schneider},
  {Sinnis}, {Smith}, {Sparks Woodle}, {Springer}, {Taboada}, {Toale},
  {Tollefson}, {Torres}, {Ukwatta}, {Villase{\~n}or}, {Weisgarber},
  {Westerhoff}, {Wisher}, {Wood}, {Yodh}, {Younk}, {Zaborov}, {Zepeda}, {Zhou},
  and {HAWC Collaboration}}]{Abeysekara-etal-2014}
{Abeysekara}, A.~U., {Alfaro}, R., {Alvarez}, C., {{\'A}lvarez}, J.~D.,
  {Arceo}, R., {Arteaga-Vel{\'a}zquez}, J.~C., {Ayala Solares}, H.~A.,
  {Barber}, A.~S., {Baughman}, B.~M., {Bautista-Elivar}, N., {Belmont}, E.,
  {BenZvi}, S.~Y., {Berley}, D., {Bonilla Rosales}, M., {Braun}, J.,
  {Caballero-Mora}, K.~S., {Carrami{\~n}ana}, A., {Castillo}, M., {Cotti}, U.,
  {Cotzomi}, J., {de la Fuente}, E., {De Le{\'o}n}, C., {DeYoung}, T., {Diaz
  Hernandez}, R., {D{\'{\i}}az-V{\'e}lez}, J.~C., {Dingus}, B.~L., {DuVernois},
  M.~A., {Ellsworth}, R.~W., {Fiorino}, D.~W., {Fraija}, N., {Galindo}, A.,
  {Garfias}, F., {Gonz{\'a}lez}, M.~M., {Goodman}, J.~A., {Gussert}, M.,
  {Hampel-Arias}, Z., {Harding}, J.~P., {H{\"u}ntemeyer}, P., {Hui}, C.~M.,
  {Imran}, A., {Iriarte}, A., {Karn}, P., {Kieda}, D., {Kunde}, G.~J., {Lara},
  A., {Lauer}, R.~J., {Lee}, W.~H., {Lennarz}, D., {Le{\'o}n Vargas}, H.,
  {Linnemann}, J.~T., {Longo}, M., {Luna-Garc{\'{\i}}a}, R., {Malone}, K.,
  {Marinelli}, A., {Marinelli}, S.~S., {Martinez}, H., {Martinez}, O.,
  {Mart{\'{\i}}nez-Castro}, J., {Matthews}, J.~A.~J., {McEnery}, J., {Mendoza
  Torres}, E., {Miranda-Romagnoli}, P., {Moreno}, E., {Mostaf{\'a}}, M.,
  {Nellen}, L., {Newbold}, M., {Noriega-Papaqui}, R., {Oceguera-Becerra}, T.,
  {Patricelli}, B., {Pelayo}, R., {P{\'e}rez-P{\'e}rez}, E.~G., {Pretz}, J.,
  {Rivi{\`e}re}, C., {Rosa-Gonz{\'a}lez}, D., {Ruiz-Velasco}, E., {Ryan}, J.,
  {Salazar}, H., {Salesa Greus}, F., {Sandoval}, A., {Schneider}, M., {Sinnis},
  G., {Smith}, A.~J., {Sparks Woodle}, K., {Springer}, R.~W., {Taboada}, I.,
  {Toale}, P.~A., {Tollefson}, K., {Torres}, I., {Ukwatta}, T.~N.,
  {Villase{\~n}or}, L., {Weisgarber}, T., {Westerhoff}, S., {Wisher}, I.~G.,
  {Wood}, J., {Yodh}, G.~B., {Younk}, P.~W., {Zaborov}, D., {Zepeda}, A.,
  {Zhou}, H., {HAWC Collaboration}, Dec. 2014. {Observation of Small-scale
  Anisotropy in the Arrival Direction Distribution of TeV Cosmic Rays with
  HAWC}. \apj 796, 108.

\bibitem[{{Ahlers} and {Mertsch}(2015)}]{Ahlers-Mertsch-2015}
{Ahlers}, M., {Mertsch}, P., Dec. 2015. {Small-scale Anisotropies of Cosmic
  Rays from Relative Diffusion}. \apjl 815, L2.

\bibitem[{{Amenomori} et~al.(2011){Amenomori}, {Bi}, {Chen}, {Cui},
  {Danzengluobu}, {Ding}, {Ding}, {Fan}, {Feng}, {Feng}, {Feng}, {Gao}, {Geng},
  {Gou}, {Guo}, {He}, {He}, {Hibino}, {Hotta}, {Hu}, {Hu}, {Huang}, {Huang},
  {Jia}, {Jiang}, {Kajino}, {Kasahara}, {Katayose}, {Kato}, {Kawata},
  {Labaciren}, {Le}, {Li}, {Li}, {Li}, {Liu}, {Lou}, {Lu}, {Meng}, {Mizutani},
  {Mu}, {Munakata}, {Nanjo}, {Nishizawa}, {Ohnishi}, {Ohta}, {Ozawa}, {Saito},
  {Saito}, {Sakata}, {Sako}, {Shibata}, {Shiomi}, {Shirai}, {Sugimoto},
  {Takita}, {Tan}, {Tateyama}, {Torii}, {Tsuchiya}, {Udo}, {Wang}, {Wang},
  {Wang}, {Wang}, {Wu}, {Xue}, {Yamamoto}, {Yan}, {Yang}, {Yasue}, {Ye}, {Yu},
  {Yuan}, {Yuda}, {Zhang}, {Zhang}, {Zhang}, {Zhang}, {Zhang}, {Zhang},
  {Zhang}, {Zhaxisangzhu}, and {Zhou}}]{Amenomori-etal-2011}
{Amenomori}, M., {Bi}, X.~J., {Chen}, D., {Cui}, S.~W., {Danzengluobu}, {Ding},
  L.~K., {Ding}, X.~H., {Fan}, C., {Feng}, C.~F., {Feng}, Z., {Feng}, Z.~Y.,
  {Gao}, X.~Y., {Geng}, Q.~X., {Gou}, Q.~B., {Guo}, H.~W., {He}, H.~H., {He},
  M., {Hibino}, K., {Hotta}, N., {Hu}, H., {Hu}, H.~B., {Huang}, J., {Huang},
  Q., {Jia}, H.~Y., {Jiang}, L., {Kajino}, F., {Kasahara}, K., {Katayose}, Y.,
  {Kato}, C., {Kawata}, K., {Labaciren}, {Le}, G.~M., {Li}, A.~F., {Li}, H.~C.,
  {Li}, J.~Y., {Liu}, C., {Lou}, Y., {Lu}, H., {Meng}, X.~R., {Mizutani}, K.,
  {Mu}, J., {Munakata}, K., {Nanjo}, H., {Nishizawa}, M., {Ohnishi}, M.,
  {Ohta}, I., {Ozawa}, S., {Saito}, T., {Saito}, T.~Y., {Sakata}, M., {Sako},
  T.~K., {Shibata}, M., {Shiomi}, A., {Shirai}, T., {Sugimoto}, H., {Takita},
  M., {Tan}, Y.~H., {Tateyama}, N., {Torii}, S., {Tsuchiya}, H., {Udo}, S.,
  {Wang}, B., {Wang}, H., {Wang}, Y., {Wang}, Y.~G., {Wu}, H.~R., {Xue}, L.,
  {Yamamoto}, Y., {Yan}, C.~T., {Yang}, X.~C., {Yasue}, S., {Ye}, Z.~H., {Yu},
  G.~C., {Yuan}, A.~F., {Yuda}, T., {Zhang}, H.~M., {Zhang}, J.~L., {Zhang},
  N.~J., {Zhang}, X.~Y., {Zhang}, Y., {Zhang}, Y., {Zhang}, Y., {Zhaxisangzhu},
  {Zhou}, X.~X., Jan. 2011. {Cosmic-ray energy spectrum around the knee
  observed with the Tibet air-shower experiment}. ASTRA 7, 15--20.

\bibitem[{{Battaner} et~al.(2015){Battaner}, {Castellano}, and
  {Masip}}]{Battaner-etal-2015}
{Battaner}, E., {Castellano}, J., {Masip}, M., Feb. 2015. {Magnetic Fields and
  Cosmic-Ray Anisotropies at TeV Energies}. \apj 799, 157.

\bibitem[{{BenZvi}(2014)}]{BenZvi-2014}
{BenZvi}, S., Jul. 2014. {Observations of the anisotropy of cosmic rays at
  TeV-PeV}. ASTRA Proceedings 1, 33--37.

\bibitem[{{Biermann} et~al.(2015){Biermann}, {Caramete}, {Meli}, {Nath}, {Seo},
  {de Souza}, and {Becker Tjus}}]{Biermann-etal-2015}
{Biermann}, P.~L., {Caramete}, L.~I., {Meli}, A., {Nath}, B.~N., {Seo}, E.-S.,
  {de Souza}, V., {Becker Tjus}, J., Oct. 2015. {Cosmic ray transport and
  anisotropies to high energies}. ASTRA Proceedings 2, 39--44.

\bibitem[{{Desiati}(2014)}]{Desiati-2014}
{Desiati}, P., Apr. 2014. {Observation of TeV-PeV cosmic ray anisotropy with
  IceCube, IceTop and AMANDA}. Nuclear Instruments and Methods in Physics
  Research A 742, 199--202.

\bibitem[{{Desiati} and {Lazarian}(2014)}]{Desiati-Lazarian-2014}
{Desiati}, P., {Lazarian}, A., Oct. 2014. {TeV Cosmic Ray Anisotropy and the
  Heliospheric Magnetic Field}. ASTRA Proceedings 1, 65--71.

\bibitem[{{Di Sciascio}(2015)}]{diSciascio-2015}
{Di Sciascio}, G., Sep. 2015. {The cosmic ray anisotropy below 10$^{15}$ eV}.
  ASTRA Proceedings 2, 27--33.

\bibitem[{{Di Sciascio} and {Iuppa}(2014)}]{DiSciascio-Iuppa-2014}
{Di Sciascio}, G., {Iuppa}, R., Jul. 2014. {On the Observation of the Cosmic
  Ray Anisotropy below 10$^{15}$ eV}. ArXiv e-prints.

\bibitem[{{Effenberger} et~al.(2012){Effenberger}, {Fichtner}, {Scherer}, and
  {B{\"u}sching}}]{Effenberger-etal-2012}
{Effenberger}, F., {Fichtner}, H., {Scherer}, K., {B{\"u}sching}, I., Nov.
  2012. {Anisotropic diffusion of Galactic cosmic ray protons and their
  steady-state azimuthal distribution}. \aap 547, A120.

\bibitem[{{Giacinti} and {Sigl}(2012)}]{Giacinit-Sigl-2012}
{Giacinti}, G., {Sigl}, G., Aug. 2012. {Local Magnetic Turbulence and TeV-PeV
  Cosmic Ray Anisotropies}. Physical Review Letters 109~(7), 071101.

\bibitem[{{Glushkov} and {Pravdin}(2013)}]{Glushkov-Pravdin-2013}
{Glushkov}, A.~V., {Pravdin}, M.~I., Feb. 2013. {Variable anisotropy of cosmic
  rays with E $_{0} < 10^{18}$ eV from Yakutsk EAS array data}. Astronomy
  Letters 39, 65--71.

\bibitem[{{Harari} et~al.(2016){Harari}, {Mollerach}, and
  {Roulet}}]{Harari-etal-2016}
{Harari}, D., {Mollerach}, S., {Roulet}, E., Mar. 2016. {Angular distribution
  of cosmic rays from an individual source in a turbulent magnetic field}. \prd
  93~(6), 063002.

\bibitem[{{Haverkorn} and {Spangler}(2013)}]{Haverkorn-Spangler-2013}
{Haverkorn}, M., {Spangler}, S.~R., Oct. 2013. {Plasma Diagnostics of the
  Interstellar Medium with Radio Astronomy}. \ssr 178, 483--511.

\bibitem[{{IceCube Collaboration} et~al.(2016){IceCube Collaboration},
  {Aartsen}, {Abraham}, {Ackermann}, {Adams}, {Aguilar}, {Ahlers}, {Ahrens},
  {Altmann}, {Anderson}, and et~al.}]{Icecube-etal-2016}
{IceCube Collaboration}, {Aartsen}, M.~G., {Abraham}, K., {Ackermann}, M.,
  {Adams}, J., {Aguilar}, J.~A., {Ahlers}, M., {Ahrens}, M., {Altmann}, D.,
  {Anderson}, T., et~al., Mar. 2016. {Anisotropy in Cosmic-Ray Arrival
  Directions in the Southern Hemisphere with Six Years of Data from the IceCube
  Detector}. ArXiv e-prints.

\bibitem[{{Iuppa} and {ARGO-YBJ Collaboration}(2013)}]{Iuppa-etal-2013}
{Iuppa}, R., {ARGO-YBJ Collaboration}, Feb. 2013. {Multi-scale TeV cosmic-ray
  anisotropy observed with the ARGO-YBJ experiment}. Journal of Physics
  Conference Series 409~(1), 012039.

\bibitem[{{Iuppa} et~al.(2012){Iuppa}, {Di Sciascio}, and {ARGO-YBJ
  Collaboration}}]{Iuppa-DiSciascio-2012}
{Iuppa}, R., {Di Sciascio}, G., {ARGO-YBJ Collaboration}, Nov. 2012.
  {Cosmic-ray anisotropies observed by the ARGO-YBJ experiment}. Nuclear
  Instruments and Methods in Physics Research A 692, 160--164.

\bibitem[{{Kopp} et~al.(2012){Kopp}, {B{\"u}sching}, {Strauss}, and
  {Potgieter}}]{Kopp-etal-2012}
{Kopp}, A., {B{\"u}sching}, I., {Strauss}, R.~D., {Potgieter}, M.~S., Mar.
  2012. {A stochastic differential equation code for multidimensional
  Fokker-Planck type problems}. Computer Physics Communications 183, 530--542.

\bibitem[{{K{\'o}ta} and {Jokipii}(2014)}]{Kota-Jokipii-2014}
{K{\'o}ta}, J., {Jokipii}, J.~R., Feb. 2014. {Are Cosmic Rays Modulated beyond
  the Heliopause?} \apj 782, 24.

\bibitem[{{Kumar} and {Eichler}(2014)}]{Kumar-Eicheler-2014}
{Kumar}, R., {Eichler}, D., Apr. 2014. {Large-scale Anisotropy of TeV-band
  Cosmic Rays}. \apj 785, 129.

\bibitem[{{Linsky} and {Redfield}(2014)}]{Linsky-Redfield-2014}
{Linsky}, J.~L., {Redfield}, S., Nov. 2014. {The local ISM in three dimensions:
  kinematics, morphology and physical properties}. \apss 354, 29--34.

\bibitem[{{L{\'o}pez-Barquero} et~al.(2015){L{\'o}pez-Barquero}, {Farber},
  {Xu}, {Desiati}, and {Lazarian}}]{Lopez-Barquero-etal-2015}
{L{\'o}pez-Barquero}, V., {Farber}, R., {Xu}, S., {Desiati}, P., {Lazarian},
  A., Sep. 2015. {Cosmic Ray Small Scale Anisotropies and Local Turbulent
  Magnetic Fields}. ArXiv e-prints.

\bibitem[{{Luo} et~al.(2015){Luo}, {Zhang}, {Potgieter}, {Feng}, and
  {Pogorelov}}]{Luo-etal-2015}
{Luo}, X., {Zhang}, M., {Potgieter}, M., {Feng}, X., {Pogorelov}, N.~V., Jul.
  2015. {A Numerical Simulation of Cosmic-Ray Modulation Near the Heliopause}.
  \apj 808, 82.

\bibitem[{{Mackey} et~al.(2015){Mackey}, {Gvaramadze}, {Mohamed}, and
  {Langer}}]{Mackey-etal-2015}
{Mackey}, J., {Gvaramadze}, V.~V., {Mohamed}, S., {Langer}, N., Jan. 2015.
  {Wind bubbles within H ii regions around slowly moving stars}. \aap 573, A10.

\bibitem[{{Mertsch} and {Funk}(2015)}]{Mertsch-Funk-2015}
{Mertsch}, P., {Funk}, S., Jan. 2015. {Solution to the Cosmic Ray Anisotropy
  Problem}. Physical Review Letters 114~(2), 021101.

\bibitem[{Parker(1965)}]{Parker-1965}
Parker, E.~N., 1965. The passage of energetic charged particles through
  interplanetary space. Planet. Space Sci. 13, 9--49.

\bibitem[{{Pierre Auger Collaboration} et~al.(2013){Pierre Auger
  Collaboration}, {Abreu}, {Aglietta}, {Ahlers}, {Ahn}, {Albuquerque},
  {Allard}, {Allekotte}, {Allen}, {Allison}, and et~al.}]{Abreu-etal-2013}
{Pierre Auger Collaboration}, {Abreu}, P., {Aglietta}, M., {Ahlers}, M., {Ahn},
  E.~J., {Albuquerque}, I.~F.~M., {Allard}, D., {Allekotte}, I., {Allen}, J.,
  {Allison}, P., et~al., Jan. 2013. {Constraints on the Origin of Cosmic Rays
  above 10$^{18}$ eV from Large-scale Anisotropy Searches in Data of the Pierre
  Auger Observatory}. \apjl 762, L13.

\bibitem[{{Pierre Auger Collaboration} et~al.(2011){Pierre Auger
  Collaboration}, {Abreu}, {Aglietta}, {Ahn}, {Albuquerque}, {Allard},
  {Allekotte}, {Allen}, {Allison}, {Alvarez Castillo}, and
  et~al.}]{Abreu-etal-2011}
{Pierre Auger Collaboration}, {Abreu}, P., {Aglietta}, M., {Ahn}, E.~J.,
  {Albuquerque}, I.~F.~M., {Allard}, D., {Allekotte}, I., {Allen}, J.,
  {Allison}, P., {Alvarez Castillo}, J., et~al., Mar. 2011. {Search for first
  harmonic modulation in the right ascension distribution of cosmic rays
  detected at the Pierre Auger Observatory}. Astroparticle Physics 34,
  627--639.

\bibitem[{{Pogorelov} et~al.(2015){Pogorelov}, {Borovikov}, {Heerikhuisen}, and
  {Zhang}}]{Pogorelov-etal-2015}
{Pogorelov}, N.~V., {Borovikov}, S.~N., {Heerikhuisen}, J., {Zhang}, M., Oct.
  2015. {The Heliotail}. \apjl 812, L6.

\bibitem[{{Potgieter}(2014)}]{Potgieter-2014}
{Potgieter}, M., Oct. 2014. {Very Local Interstellar Spectra for Galactic
  Electrons, Protons and Helium}. Brazilian Journal of Physics 44, 581--588.

\bibitem[{{Scherer} et~al.(2016){Scherer}, {Fichtner}, {Kleimann},
  {Wiengarten}, {Bomans}, and {Weis}}]{Scherer-etal-2016}
{Scherer}, K., {Fichtner}, H., {Kleimann}, J., {Wiengarten}, T., {Bomans},
  D.~J., {Weis}, K., Feb. 2016. {Shock structures of astrospheres}. \aap 586,
  A111.

\bibitem[{{Scherer} et~al.(2011){Scherer}, {Fichtner}, {Strauss}, {Ferreira},
  {Potgieter}, and {Fahr}}]{Scherer-etal-2011}
{Scherer}, K., {Fichtner}, H., {Strauss}, R.~D., {Ferreira}, S.~E.~S.,
  {Potgieter}, M.~S., {Fahr}, H.-J., Jul. 2011. {On Cosmic Ray Modulation
  beyond the Heliopause: Where is the Modulation Boundary?} \apj 735, 128--+.

\bibitem[{{Scherer} et~al.(2015){Scherer}, {van der Schyff}, {Bomans},
  {Ferreira}, {Fichtner}, {Kleimann}, {Strauss}, {Weis}, {Wiengarten}, and
  {Wodzinski}}]{Scherer-etal-2015a}
{Scherer}, K., {van der Schyff}, A., {Bomans}, D.~J., {Ferreira}, S.~E.~S.,
  {Fichtner}, H., {Kleimann}, J., {Strauss}, R.~D., {Weis}, K., {Wiengarten},
  T., {Wodzinski}, T., Apr. 2015. {Cosmic rays in astrospheres}. \aap 576, A97.

\bibitem[{{Schwadron} et~al.(2015){Schwadron}, {Adams}, {Christian}, {Desiati},
  {Frisch}, {Funsten}, {Jokipii}, {McComas}, {Moebius}, and
  {Zank}}]{Schawdron-etal-2015}
{Schwadron}, N.~A., {Adams}, F.~C., {Christian}, E., {Desiati}, P., {Frisch},
  P., {Funsten}, H.~O., {Jokipii}, J.~R., {McComas}, D.~J., {Moebius}, E.,
  {Zank}, G.~P., Jan. 2015. {Anisotropies in TeV Cosmic Rays Related to the
  Local Interstellar Magnetic Field from the IBEX Ribbon}. Journal of Physics
  Conference Series 577~(1), 012023.

\bibitem[{{Stanimirovi{\'c}} et~al.(2010){Stanimirovi{\'c}}, {Weisberg}, {Pei},
  {Tuttle}, and {Green}}]{Stanimirocic-etal-2010}
{Stanimirovi{\'c}}, S., {Weisberg}, J.~M., {Pei}, Z., {Tuttle}, K., {Green},
  J.~T., Sep. 2010. {Arecibo Multi-epoch H I Absorption Measurements Against
  Pulsars: Tiny-scale Atomic Structure}. \apj 720, 415--434.

\bibitem[{{Strauss} et~al.(2011{\natexlab{a}}){Strauss}, {Potgieter},
  {B{\"u}sching}, and {Kopp}}]{Strauss-etal-2011b}
{Strauss}, R.~D., {Potgieter}, M.~S., {B{\"u}sching}, I., {Kopp}, A., Jul.
  2011{\natexlab{a}}. {Modeling the Modulation of Galactic and Jovian Electrons
  by Stochastic Processes}. \apj 735, 83.

\bibitem[{{Strauss} et~al.(2013){Strauss}, {Potgieter}, {Ferreira}, {Fichtner},
  and {Scherer}}]{Strauss-etal-2013}
{Strauss}, R.~D., {Potgieter}, M.~S., {Ferreira}, S.~E.~S., {Fichtner}, H.,
  {Scherer}, K., Mar. 2013. {Cosmic Ray Modulation Beyond the Heliopause: A
  Hybrid Modeling Approach}. \apjl 765, L18.

\bibitem[{{Strauss} et~al.(2011{\natexlab{b}}){Strauss}, {Potgieter}, {Kopp},
  and {B{\"u}sching}}]{Strauss-etal-2011a}
{Strauss}, R.~D., {Potgieter}, M.~S., {Kopp}, A., {B{\"u}sching}, I., Dec.
  2011{\natexlab{b}}. {On the propagation times and energy losses of cosmic
  rays in the heliosphere}. \jgr 116, 12105.

\bibitem[{{Toal{\'a}} and {Arthur}(2011)}]{Toala-Arthur-2011}
{Toal{\'a}}, J.~A., {Arthur}, S.~J., Aug. 2011. {Radiation-hydrodynamic Models
  of the Evolving Circumstellar Medium around Massive Stars}. \apj 737, 100.

\bibitem[{{Toscano} and {IceCube Collaboration}(2012)}]{Toscano-etal-2012}
{Toscano}, S., {IceCube Collaboration}, Nov. 2012. {Observation of anisotropy
  in the arrival direction distribution of cosmic rays above TeV energies with
  IceCube}. Nuclear Instruments and Methods in Physics Research A 692,
  165--169.

\bibitem[{{Zhang} et~al.(2014){Zhang}, {Zuo}, and
  {Pogorelov}}]{Zhang-etal-2014}
{Zhang}, M., {Zuo}, P., {Pogorelov}, N., Jul. 2014. {Heliospheric Influence on
  the Anisotropy of TeV Cosmic Rays}. \apj 790, 5.

\bibitem[{{Zotov} and {Kulikov}(2012)}]{Zotov-Kulikov-2012}
{Zotov}, M.~Y., {Kulikov}, G.~V., Nov. 2012. {A search for small-scale
  anisotropy of PeV cosmic rays}. Astronomy Letters 38, 731--743.

\end{thebibliography}

\end{document}